\documentstyle[amssymb,thmsa,a4,sw20lart]{article}

\input{tcilatex}

\begin{document}

\author{S. Manoff$^1$, B. Dimitrov$^2$ \\
{\it 1. Institute for Nuclear Research and Nuclear Energy,}\\
{\it Department of Theoretical Physics,}\\
{\it 1784 Sofia - Bulgaria}\\
{\it 2.} {\it Joint Institute for Nuclear Research,}\\
{\it Bogoliubov Laboratory of Theoretical Physics,}\\
{\it Dubna, Moscow Region}\\
{\it 141980 Russia}}
\title{{\sc Flows and particles with shear-free and expansion-free velocities in (}$%
\overline{L}_n${\sc ,}$g${\sc )- and Weyl's spaces}}
\date{e-mail addresses: {\it \ smanov@inrne.bas.bg, bogdan@thsun1.jinr.ru}}
\maketitle

\begin{abstract}
Conditions for the existence of flows with non-null shear-free and
expansion-free velocities in spaces with affine connections and metrics are
found. On their basis, generalized Weyl's spaces with shear-free and
expansion-free conformal Killing vectors as velocities vectors of spinless
test particles moving in a Weyl's space \ are considered. The necessary and
sufficient conditions are found under which a free spinless test particle
could move in spaces with affine connections and metrics on a curve
described by means of an auto-parallel equation. In Weyl's spaces with
Weyl's covector, constructed by the use of a dilaton field, the dilaton
field appears as a scaling factor for the rest mass density of the test
particle.

\end{abstract}

\section{Introduction}

\subsection{Weyl's spaces as special case of spaces with affine connections
and metrics}

In the last years Weyl's spaces have deserved some interest related to the
possibility of using mathematical models of space-time different from
(pseudo) Riemannian spaces without torsion ($V_n$-spaces) or with torsion ($%
U_n$-spaces)\cite{Hehl-1} $\div $ \cite{Salim}. On the one side, Weyl's
spaces appear as a generalization of $V_n$- and $U_n$-spaces. On the other
side, they are special cases of spaces with affine connections and metrics.
The use of spaces with affine connections and metrics as models of
space-time has been critically evaluated from different points of view \cite
{Hayashi}, \cite{Treder}. But recently, it has been proved that in spaces
with contravariant and covariant affine connections (whose components differ
only by sign) and metrics [$(L_n,g)$-spaces] as well as in spaces with
contravariant and covariant affine connections (whose components differ not
only by sign) and metrics [$(\overline{L}_n,g)$-spaces] \cite{Manoff-1} \cite
{Manoff-1a}

\begin{itemize}
\item  the principle of equivalence holds \cite{Iliev-1}$\div $\cite{Hartley}%
,

\item  special types of transports (called Fermi-Walker transports) \cite
{Manoff-2} $\div $ \cite{Manoff-3} exist which do not deform a Lorentz
basis, and therefore, the low of causality is not abused in $(L_n,g)$- and $(%
\overline{L}_n,g)$-spaces if one uses a Fermi-Walker transport instead of a
parallel transport (used in a $V_n$-space).

\item  there also exist other types of transports (called conformal
transports) \cite{Manoff-4}, \cite{Manoff-5} under which a light cone does
not deform,

\item  the auto-parallel equation can play the same role in $(L_n,g)$- and $(%
\overline{L}_n,g)$-spaces as the geodesic equation does in the Einstein
theory of gravitation (ETG) in $V_n$-spaces \cite{Manoff-6}, \cite{Manoff-7}%
, where the geodesic equation is identical with the auto-parallel equation.
\end{itemize}

On this basis, many of the differential-geometric constructions used in the
ETG in $V_4$-spaces could be generalized for the cases of $(L_n,g)$- and $(%
\overline{L}_n,g)$-spaces, and especially for Weyl's spaces without torsion (%
$W_n$ or $\overline{W}_n$-spaces) or in Weyl's spaces with torsion ($Y_n$-
or $\overline{Y}_n$-spaces) as special cases of $(L_n,g)$- or $(\overline{L}%
_n,g)$-spaces. Bearing in mind this background, a question arises about
possible physical applications and interpretation of mathematical
constructions from ETG generalized for Weyl's spaces with affine connections
and metrics. On the other side, it is well known that every classical field
theory over spaces with affine connections and metrics could be considered
as a theory of continuous media in these spaces \cite{Hehl-3} $\div $ \cite
{Manoff-8b}. On this ground, notions of the continuous media mechanics (such
as deformation velocity and acceleration, shear velocity and acceleration,
rotation velocity and acceleration, expansion velocity and acceleration) 
\cite{Manoff-preprint} have been used as invariant characteristics for
spaces admitting vector fields with special kinematic characteristics \cite
{Ehlers}, \cite{Stephani}.

The existence of a flow with shear-free and expansion-free velocity is of
great importance for continuous media mechanics in the relativistic and
non-relativistic case as well as for hydrodynamics as a part of it \cite
{Manoff-preprint}. Such type of a flow exists when the flow does not change
its form and volume, i.e. it does not deform, expand or shrink (contract).
Flows with these properties are considered as stable during their motion in
the space or in the space-time.

{\it The main task of this paper \ is the investigation of Weyl's spaces
with respect to their ability to admit flows and spinless test particles
with velocities that are shear-free and expansion-free vector fields}. On
this basis, conditions for the existence of shear-free and expansion-free
non-null (non-isotropic) vector fields in spaces with affine connections and
metrics are found and then specialized for Weyl's spaces. At the same time,
a possible interpretation of a dilaton field, appearing in the structure of
special types of Weyl's spaces, is found on the basis of the auto-parallel
equation describing the motion of a free spinless test particle in these
types of spaces.

In Section 2, the notions of relative velocity, shear velocity, and
expansion velocity are considered. The equivalence of the action of the Lie
differential operator and of the covariant differential operator on the
invariant volume element generates restrictions to a vector field along
which these operators act and respectively to the existing in the manifold
metrics. It is shown that the equivalence condition appears as a condition
for the existence of shear-free and expansion-free non-null vector fields as
velocities vector fields of flows or particles in spaces with affine
connections and metrics. The same condition in a Weyl's space appears as a
condition for the existence of a shear-free and expansion-free conformal
Killing vector field of special type. Sufficient conditions are found under
which in special type of Weyl's spaces non-null auto-parallel, shear-free,
and expansion-free conformal Killing vector fields could exist as velocities
vector fields of a flow or of spinless test particles. In Section 3 the
auto-parallel equation in Weyl's spaces is discussed as an equation for
describing the motion of a free moving spinless test particle. Concluding
remarks comprise the final Section 4. The most considerations are given in
details (even in full details) for those readers who are not familiar with
the investigated problems.

\subsection{Abbreviations, definitions, and symbols}

In the further considerations in this paper we will use the following
abbreviations, definitions and symbols:

:= means by definition.

Point '' $\cdot $'' is used as a symbol for standard multiplication in the
field of real (or complex) numbers, e.g. $a\cdot b\in {\bf R}$.

The point '' $.$'' is used as a symbol for symmetric tensor product, e.g. $%
u.v=\frac 12\cdot (u\otimes v+v\otimes u)$.

''$\wedge $'' is the symbol for a wedge product, e.g. $u\wedge v=\frac
12\cdot (u\otimes v-v\otimes u)$.

$M$ is a symbol for a differentiable manifold with $\dim M=n$. $T(M):=\cup
_{x\in M}T_{x}(M)$ and $T^{\ast }(M):=\cup _{x\in M}T_{x}^{\ast }(M)$ are
the tangent and the cotangent spaces at $M$ respectively.

$(\overline{L}_n,g)$, $\overline{Y}_n$, $\overline{U}_n$, and $\overline{V}%
_n $ are spaces with contravariant and covariant affine connections and
metrics whose components {\it differ not only by sign } \cite{Manoff-1}. In
such type of spaces the non-canonical contraction operator $S$ acts on a
contravariant basic vector field $e_j$ (or $\partial _j$)$\,\in \{e_j$ (or $%
\partial _j$)$\}\subset T(M)$ and on a covariant \ basic vector field $e^i\,$%
\ (or $dx^i$) $\in \{e^i$ (or $dx^i$)$\}\subset T^{*}(M)$ in the form 
\begin{eqnarray*}
S &:&(e^i,e_j)\longrightarrow S(e^i,e_j):=S(e_j,e^i):=f^i\,_j\text{ , \ \ \
\ \ \ } \\
\text{\ \ \ \ }f^i\,_j &\in &C^r(M)\text{ , \ \ \ \ \ \ \ \ }r\geqq 2\text{
, \ \ \ \ \ \ }\det (f^i\,_j)\neq 0\text{, \ \ \ \ \ \ } \\
\exists \text{\ \ \ }f_i\,^k &\in &C^r(M)\text{, \ \ \ \ \ \ \ \ \ }r\geqq 2%
\text{ }:\text{\ \ \ \ \ \ \ \ }f^i\,_j\cdot f_i\,^k:=g_j^k\text{.\ \ \ \ \
\ \ \ \ }
\end{eqnarray*}

The functions $f^i\,_j=f^i\,_j(x^k)$ fulfil equations 
\[
f^i\,_{j,k}=\Gamma _{jk}^l\cdot f^i\,_l+P_{lk}^i\cdot f^l\,_j\text{
\thinspace \thinspace ,\thinspace \thinspace \thinspace \thinspace
\thinspace \thinspace \thinspace \thinspace \thinspace \thinspace \thinspace
\thinspace \thinspace \thinspace \thinspace \thinspace \thinspace \thinspace 
}f^i\,_{j,k}=\partial _kf^i\,_j=\frac{\partial f^i\,_j}{\partial x^k}\text{ ,%
} 
\]

\noindent related to the both affine connections $\Gamma $ and $P$ with
components $\Gamma _{jk}^i$ and $P_{jk}^i$ in the co-ordinate bases $%
\{\partial _i\}$ and $\{dx^i\}$ respectively.

In these spaces, for example, $g(u)=g_{ik}\cdot f^{k}\,_{j}\cdot u^{j}\cdot
dx^{i}:=g_{i\overline{j}}\cdot u^{j}\cdot dx^{i}=g_{ij}\cdot u^{\overline{j}%
}\cdot dx^{i}:=u_{i}\cdot dx^{i}$, $g(u,u)=g_{kl}\cdot f^{k}\,_{i}\cdot
f^{l}\,_{j}\cdot u^{i}\cdot u^{j}:=g_{\overline{i}\overline{j}}\cdot
u^{i}\cdot u^{j}=g_{ij}\cdot u^{\overline{i}}\cdot u^{\overline{j}%
}=u_{j}\cdot u^{\overline{j}}:=u_{\overline{i}}\cdot u^{i}$, $g^{\overline{i}%
\overline{j}}\cdot g_{jk}=\delta _{k}^{i}=g_{k}^{i}$, $g_{\overline{i}%
\overline{k}}.g^{kj}=g_{i}^{j}$. The components $\delta _{j}^{i}:=g_{j}^{i}$
($\mid =0$ for $i\neq j$ and $\mid =1$ for $i=j$) are the components of the
Kronecker tensor $Kr:=g_{j}^{i}\cdot \partial _{i}\otimes dx^{j}$.

$(L_n,g)$, $Y_n$, $U_n$, and $V_n$ are spaces with contravariant and
covariant affine connections and metrics whose components {\it differ only
by sign} \cite{Norden}, \cite{Raschewski}. In such type of spaces the
canonical contraction operator $S:=C$ acts on a contravariant basic vector
field $e_j$ (as a non-co-ordinate, non-holonomic contravariant basic vector
field) [or $\partial _j$ (as a co-ordinate, holonomic contravariant basic
vector field)]$\,\in \{e_j$ (or $\partial _j$)$\}\subset T(M)$ and on a
covariant \ basic vector field $e^i$\thinspace (as a non-co-ordinate,
non-holonomic covariant basic vector field)\ [or $dx^i$ as a co-ordinate,
holonomic covariant basic vector field)] $\in \{e^i$ (or $dx^i$)$\}\subset
T^{*}(M)$ in the form 
\[
C:(e^i,e_j)\longrightarrow C(e^i,e_j):=C(e_j,e^i):=\delta _j^i:=g_j^i\text{ .%
} 
\]

In these spaces, for example, $g(u)=g_{ik}\cdot g_j^k\cdot u^j\cdot
dx^i:=g_{ij}\cdot u^j\cdot dx^i=u_i\cdot dx^i$, $g(u,u)=g_{kl}\cdot
g_i^k\cdot g_j^l\cdot u^i\cdot u^j:=g_{ij}\cdot u^i\cdot u^j=u_i\cdot u^i$.

The different types of spaces with affine connections and metrics with
respect to their properties related to the contraction operator $S$ (or $C$)
and to the metric $g$ could be given roughly in the following scheme

$
\begin{array}{cccc}
\text{{\bf Space}} & 
\begin{array}{c}
\text{Contraction} \\ 
\text{operator}
\end{array}
& 
\begin{array}{c}
\text{Components} \\ 
\text{of the contravariant} \\ 
\text{and covariant} \\ 
\text{affine connections }\Gamma \text{ and }P
\end{array}
& 
\begin{array}{c}
\text{Covariant} \\ 
\text{derivative of} \\ 
\text{the metric with} \\ 
\text{respect to} \\ 
\text{the affine connections}
\end{array}
\\ 
&  &  &  \\ 
(\overline{L}_n,g)\text{-space} & S=f^i\,_j\cdot \partial _i\otimes dx^j & 
\begin{array}{c}
\Gamma _{jk}^i\neq -P_{jk}^i \\ 
\Gamma _{jk}^i\neq \Gamma _{kj}^i \\ 
P_{jk}^i\neq P_{kj}^i
\end{array}
& g_{ij;k}\neq 0 \\ 
(L_n,g)\text{-space} & S=C=g_j^i\cdot \partial _i\otimes dx^j & 
\begin{array}{c}
\Gamma _{jk}^i=-P_{jk}^i \\ 
\Gamma _{jk}^i\neq \Gamma _{kj}^i
\end{array}
& g_{ij;k}\neq 0 \\ 
\begin{array}{c}
\overline{Y}_n\text{-space} \\ 
\text{(Weyl}^{\prime }\text{s space} \\ 
\text{with torsion)}
\end{array}
& S=f^i\,_j\cdot \partial _i\otimes dx^j & 
\begin{array}{c}
\Gamma _{jk}^i\neq -P_{jk}^i \\ 
\Gamma _{jk}^i\neq \Gamma _{kj}^i \\ 
P_{jk}^i\neq P_{kj}^i
\end{array}
& g_{ij;k}=\frac 1n\cdot Q_k\cdot g_{ij} \\ 
\begin{array}{c}
Y_n\text{-space} \\ 
\text{(Weyl}^{\prime }\text{s space} \\ 
\text{with torsion)}
\end{array}
& S=C=g_j^i\cdot \partial _i\otimes dx^j & 
\begin{array}{c}
\Gamma _{jk}^i=-P_{jk}^i \\ 
\Gamma _{jk}^i\neq \Gamma _{kj}^i
\end{array}
& g_{ij;k}=\frac 1n\cdot Q_k\cdot g_{ij} \\ 
\begin{array}{c}
\overline{U}_n\text{-space} \\ 
\text{(Pseudo-} \\ 
\text{Riemannian} \\ 
\text{space with} \\ 
\text{torsion)}
\end{array}
& S=f^i\,_j\cdot \partial _i\otimes dx^j & 
\begin{array}{c}
\Gamma _{jk}^i\neq -P_{jk}^i \\ 
\Gamma _{jk}^i\neq \Gamma _{kj}^i \\ 
P_{jk}^i\neq P_{kj}^i
\end{array}
& g_{ij;k}=0 \\ 
\begin{array}{c}
U_n\text{-space} \\ 
\text{(Pseudo-} \\ 
\text{Riemannian} \\ 
\text{space with} \\ 
\text{torsion)}
\end{array}
& S=C=g_j^i\cdot \partial _i\otimes dx^j & 
\begin{array}{c}
\Gamma _{jk}^i=-P_{jk}^i \\ 
\Gamma _{jk}^i\neq \Gamma _{kj}^i
\end{array}
& g_{ij;k}=0 \\ 
\begin{array}{c}
\overline{V}_n\text{-space} \\ 
\text{(Pseudo-} \\ 
\text{Riemannian} \\ 
\text{space without} \\ 
\text{torsion)}
\end{array}
& S=f^i\,_j\cdot \partial _i\otimes dx^j & 
\begin{array}{c}
\Gamma _{jk}^i\neq -P_{jk}^i \\ 
\Gamma _{jk}^i\neq \Gamma _{kj}^i \\ 
P_{jk}^i=P_{kj}^i
\end{array}
& g_{ij;k}=0 \\ 
\begin{array}{c}
V_n\text{-space} \\ 
\text{(Pseudo-} \\ 
\text{Riemannian} \\ 
\text{space without} \\ 
\text{torsion)}
\end{array}
& S=C=g_j^i\cdot \partial _i\otimes dx^j & 
\begin{array}{c}
\Gamma _{jk}^i=-P_{jk}^i \\ 
\Gamma _{jk}^i=\Gamma _{kj}^i
\end{array}
& g_{ij;k}=0
\end{array}
$

{\it Remark}. All results found for $(\overline{L}_{n},g)$-spaces could be
specialized for $(L_{n},g)$-spaces by omitting the bars above or under the
indices.

$\nabla _{u}$ is the covariant differential operator acting on the elements
of the tensor algebra ${\cal T}$ over $M$. The action of \ $\nabla _{u}$ is
called covariant differentiation (covariant transport) along a contravariant
vector field $u$, for instance, 
\begin{equation}
\nabla _{u}v:=v_{\;;j}^{i}\cdot u^{j}\cdot \partial
_{i}=(v^{i}\,_{,j}+\Gamma _{kj}^{i}\cdot v^{k})\cdot u^{j}\cdot \partial _{i}%
\text{ , \ \ \ \ \ }v\in T(M)\text{ ,}  \label{0.1}
\end{equation}
where $v^{i}\,_{,j}:=\partial v^{i}/\partial x^{j}$ and $\Gamma _{jk}^{i}$
are the components of the contravariant affine connection $\Gamma $ in a
contravariant co-ordinate basis $\{\partial _{i}\}$. The result $\nabla
_{u}v $ of the action of $\nabla _{u}$ on a tensor field $v\in \otimes
_{l}^{k}(M)$ is called covariant derivative of $v$ along $u$. For covariant
vectors and tensor fields an analogous relation holds, for instance, 
\begin{equation}
\nabla _{u}w=w_{i;j}\cdot u^{j}\cdot dx^{i}=(w_{i,j}+P_{ij}^{l}.w_{l})\cdot
u^{j}\cdot .dx^{i}\text{ \ , \ \ \ }w\in T^{\ast }(M)\text{ .}  \label{0.2}
\end{equation}
where $P_{ij}^{l}$ are the components of the covariant affine connection $P$
in a covariant co-ordinate basis $\{dx^{i}\}$. For $(L_{n},g)$, $Y_{n}$, $%
W_{n}$, $U_{n}$, and $V_{n}$-spaces $P_{ij}^{l}=-\Gamma _{ij}^{l}$.

$\pounds _u$ is the Lie differential operator \cite{Manoff-1} acting on the
elements of the tensor algebra ${\cal T}$ over $M$. The action of \ $\pounds
_u$ is called dragging-along a contravariant vector field $u$. The result $%
\pounds _uv$ of the action of $\pounds _u$ on a tensor field $v$ is called
Lie derivative of $v$ along $u$. The Lie derivative of a contravariant
vector $\xi $ along a contravariant vector $u$ could be written in the form 
\[
\pounds _u\xi =\nabla _u\xi -\nabla _\xi u-T(u,\xi )\text{ \thinspace .} 
\]

The contravariant vector $T(u,\xi )$ is called contravariant torsion vector.
The tensor $T=T_{ij}\,^k\cdot dx^i\wedge dx^j\otimes \partial _k$ is called
contravariant torsion tensor. Its components $T_{ij}\,^k$ in a co-ordinate
basis have the form 
\[
T_{ij}\,^k:=\Gamma _{ji}^k-\Gamma _{ij}^k\text{ \thinspace \thinspace .} 
\]

The $n$-form $d\omega :=\frac 1{n!}\cdot \sqrt{-d_g}\cdot \varepsilon
_{i_1...i_n}\cdot dx^{i_1}\wedge ...\wedge x^{i_n}$, where $d_g:=\det
(g_{ij})<0$, $\varepsilon _{i_1...i_n}$ are the components of the full
antisymmetric Levi-Civita symbol, is called invariant volume element in $M$.

{\it Line of a flow} is a line with tangent vector at each of its points
collinear (or identical) with the velocity of a material point with the
corresponding position \cite{Manoff-preprint}.

{\it Trajectory of a particle.} A line of a flow could be considered
separately as the trajectory of a particle moving in a $(\overline{L}_n,g)$%
-space.

{\it Flow} is a congruence (family) of flow's lines.

For the further considerations we need to recall some properties of the
Weyl's spaces and the conformal Killing vector fields.

\subsection{Properties of a Weyl's space with affine connections and metrics}

Usually, a Weyl's space without torsion and with contraction operator $S=C$
is defined by means of the condition for the vanishing of the covariant
derivative of a metric $\overline{g}:=\varphi \cdot g$ conformal to a given
metric $g$, where $\varphi \in C^r(M)$, 
\[
\nabla _u\overline{g}=\nabla _u(\varphi \cdot g):=0\,\,\,\,\,\text{ for
\thinspace \thinspace \thinspace \thinspace }\forall u\in T(M)\text{
\thinspace .\thinspace \thinspace } 
\]

The introduced definition is equivalent to the definition for a Weyl's space
as a space fulfilling the condition 
\[
\nabla _ug=q_u\cdot g\text{ \thinspace \thinspace \thinspace ,\thinspace
\thinspace \thinspace \thinspace \thinspace \thinspace \thinspace \thinspace
\thinspace \thinspace \thinspace \thinspace }q_u=-u(log\varphi )\text{
\thinspace \thinspace ,\thinspace \thinspace \thinspace \thinspace
\thinspace \thinspace \thinspace \thinspace \thinspace }\varphi \in C^r(M)%
\text{ \thinspace \thinspace \thinspace ,} 
\]
and leads automatically to the proposition $1$ and $2$ considered below.

A more general definition for a Weyl's space with affine connections and
metrics (which includes $\overline{Y}_n$- and $Y_n$-spaces) could be
introduced on the basis of a recurrent relation for the metric $g$.

1. Let us now consider the condition for the existence of a Weyl's space
with affine connections and metrics on the basis of the following definition

{\it Definition 1. }A Weyl's space is a differentiable manifold $M$ with $%
\dim M:=n$, provided with affine connections $\Gamma $ and $P$ (with $P\neq
-\Gamma $ or $P=-\Gamma $) and a metric $g$ with covariant derivative of $g$
along an arbitrary given contravariant vector field $u\in T(M)$ in the form 
\begin{equation}
\nabla _ug:=\frac 1n\cdot Q_u\cdot g\text{ .}  \label{1.7}
\end{equation}
The existence condition is a recurrent relation for the metric $g$. Here 
\begin{eqnarray}
Q_u &=&Q_j\cdot u^j\text{ ,\thinspace \ \ \ \ }Q:=Q_j\cdot dx^j\text{ ,\ \ \
\ }  \label{1.8} \\
\text{\ }Q_j &:&=\underline{Q}_k\cdot f^k\,_j:=\underline{Q}_{\overline{j}}%
\text{ for }P\neq -\Gamma  \nonumber \\
Q_j &\equiv &Q_j\text{ for }P=-\Gamma \text{ .}  \nonumber
\end{eqnarray}

The covariant vector field (1-form) $\overline{Q}:=\frac{1}{n}\cdot
Q_{j}\cdot dx^{j}=\frac{1}{n}\cdot Q$ is called Weyl's covariant (covector)
field. If $Q$ is an exact form, i. e. if $Q=-d\overline{\varphi }=-\overline{%
\varphi }_{,j}\cdot dx^{j}$ with $Q_{j}=-\overline{\varphi }_{,j}$, $%
\overline{\varphi }\in C^{r}(M)$, $r\geqq 2$, then for a contravariant
vector field $u:=d/d\tau $ the invariant $Q_{u}$ could be written in the
form $Q_{u}=-d\overline{\varphi }/d\tau $. The scalar field $\overline{%
\varphi }$ is called dilaton field. The reason for this notation follows
from the properties of the Weyl's spaces considered below.

After contracting $\nabla _{u}g$ and $g$ from the last equation with $%
\overline{g}=g^{ij}\cdot \partial _{i}.\partial _{j}$ by both basic vector
fields, i.e. after finding the relations 
\begin{equation}
\overline{g}[\nabla _{u}g]=g^{\overline{i}\overline{j}}\cdot g_{ij;k}\cdot
u^{k}\text{ \ \ \ \ \ \ \ and \ \ \ \ \ \ \ }\overline{g}[g]=g^{\overline{k}%
\overline{l}}.g_{kl}=n=\dim M\text{ \ ,}  \label{1.9}
\end{equation}
it follows for $Q_{u}$%
\begin{equation}
Q_{u}=\overline{g}[\nabla _{u}g]=2\cdot \text{\ }\overline{f}_{u}\text{ \ ,
\ \ \ \ \ }\overline{f}_{u}=\frac{1}{2}\cdot Q_{u}\text{ \ \ \ \ \ .}
\label{1.10}
\end{equation}

Therefore, for Weyl's spaces, we have the recurrent relation for the
invariant volume element $d\omega $%
\begin{equation}
\nabla _u(d\omega )=\frac 12\cdot Q_u.d\omega \text{ \ \ \ \ .}  \label{1.11}
\end{equation}

2. Parallel transports over Weyl's spaces are at the same time conformal
transports. This means that if $\nabla _u\xi =0$ and $\nabla _u\eta =0$,
then $ul_\xi =(1/2n)\cdot Q_u\cdot l_\xi $, $ul_\eta =(1/2n)\cdot Q_u\cdot
l_\eta $, and \ $u[\cos (\xi ,\eta )]=0$, where $l_\xi :=\mid g(\xi ,\xi
)\mid ^{1/2}$, $l_\eta :=\;\mid g(\eta ,\eta )\mid ^{1/2}$, $\cos (\xi ,\eta
):=[g(\xi ,\eta )]/(l_\xi \cdot l_\eta )$. If $u=d/d\tau $, then \cite
{Manoff-e} 
\begin{equation}
\frac{dl_\xi }{d\tau }=\frac 1{2n}\cdot Q_u\cdot l_\xi \text{ , \ \ \ \ \ \ }%
\frac{dl_\eta }{d\tau }=\frac 1{2n}\cdot Q_u\cdot l_\eta \text{\ \ \ \ ,}
\label{1.25}
\end{equation}
and therefore, 
\begin{eqnarray}
l_\xi &=&l_{\xi 0}\cdot \exp [\frac 1{2n}\cdot \int Q_j\cdot dx^j]\text{ , \ 
}l_\eta =l_{\eta 0}\cdot \exp [\frac 1{2n}\cdot \int Q_j\cdot dx^j]\text{ ,
\ \ }  \label{1.26} \\
\text{\ }l_{\xi 0} &=&\text{const., }l_{\eta 0}=\text{const., \ \ \ \ \ }%
\cos (\xi ,\eta )=\text{const. along }\tau (x^k)\text{. \ \ \ \ \ \ } 
\nonumber
\end{eqnarray}

Furthermore, if $Q_{j}=-n\cdot \overline{\varphi }_{,j}$ and respectively $%
Q_{u}=-n\cdot d\overline{\varphi }/d\tau $ , then the equation for $l_{\xi }$
obtains in this case the simple form 
\begin{equation}
\frac{dl_{\xi }}{d\tau }=-\frac{1}{2}\cdot \frac{d\overline{\varphi }}{d\tau 
}\cdot l_{\xi }\text{ .}  \label{1.27}
\end{equation}

The solution for $l_{\xi }$ could easily be found as 
\begin{equation}
l_{\xi }=l_{\xi 0}\cdot e^{-\frac{1}{2}\cdot \overline{\varphi }}\text{ \ \
\ .}  \label{1.27a}
\end{equation}

The scalar field $\overline{\varphi }$ [as an invariant function $\overline{%
\varphi }\in \otimes ^{0}\,_{0}(M)$] appears as a gauge factor changing the
length of the vector $\xi $. This is the reason for calling the scalar field 
$\overline{\varphi }$ {\it dilaton field} in a Weyl's space.

3. \ The metric in a Weyl's space has properties which can be formulated in
the following two propositions:

{\it Proposition 1}. \cite{Bonneau} A metric $\widetilde{g}$ conformal to a
Weyl's metric $g$ is also a Weyl's metric. In other words, if $\widetilde{g}%
=e^\varphi \cdot g$ with $\nabla _\xi g=\frac 1n\cdot Q_\xi \cdot g$ for $%
\forall \xi \in T(M)$, then $\ \nabla _\xi \widetilde{g}=\frac 1n\cdot 
\widetilde{Q}_\xi \cdot \widetilde{g}$.

The proof is trivial.

Therefore, all Weyl's metrics belong to the set of all metrics conformal to
a Weyl's metric.

Let the square $ds^{2}$ of the line element $ds$ for a Weyl's metric $g$ in $%
\overline{W}_{n}$- (or $\overline{Y}_{n}$)-spaces and in $W_{n}$- (or $Y_{n}$%
)-spaces be given in the forms respectively 
\[
ds^{2}=g_{\overline{i}\overline{j}}\cdot dx^{i}\cdot dx^{j}\text{ \ \ , \ \
\ \ \ \ \ \ \ \ }ds^{2}=g_{ij}\cdot dx^{i}\cdot dx^{j}\text{ \ \ .} 
\]

Then, the square $d\widetilde{s}^2$ of the line element $d\widetilde{s}$ for
a conformal to the Weyl's metric $g$ will have the forms respectively 
\[
d\widetilde{s}^2=\widetilde{g}_{\overline{i}\overline{j}}\cdot dx^i\cdot
dx^j=e^\varphi .ds^2\text{ \ \ , \ \ \ \ \ \ \ \ \ \ }d\widetilde{s}^2=%
\widetilde{g}_{ij}\cdot dx^i\cdot dx^j=e^\varphi .ds^2\text{ \ \ .} 
\]

The invariant function $\varphi =\varphi (x^{k})\in \otimes
_{0}^{0}(M)\subset C^{r}(M)$, $r\geqq 2$, is also called {\it dilaton field}
because of its appearing as a gauge factor changing the line element in a
Weyl's space.

For $\pm \widetilde{l}_{d}^{2}=\widetilde{g}(d,d)=d\widetilde{s}^{2}=%
\widetilde{g}_{\overline{i}\overline{j}}\cdot dx^{i}\cdot dx^{j}$ with $%
d^{i}:=dx^{i}$ we have 
\[
\pm \widetilde{l}_{d}^{2}=d\widetilde{s}^{2}=e^{\varphi }.ds^{2}=\pm
e^{\varphi }.l_{d}^{2}\text{\ .} 
\]

On the other side, by the use of the expression (\ref{1.26}) for the length $%
\widetilde{l}_{d}$ we find that 
\begin{eqnarray*}
\widetilde{l}_{d}^{2} &=&\widetilde{l}_{d0}^{2}.\exp [\frac{1}{n}\cdot \int 
\widetilde{Q}_{j}\cdot dx^{j}]=\widetilde{l}_{d0}^{2}\cdot \exp [\frac{1}{n}%
\cdot \int (n\cdot \varphi _{,j}+Q_{j})\cdot dx^{j}= \\
&=&\widetilde{l}_{d0}^{2}\cdot \exp \varphi \cdot \exp [\frac{1}{n}\cdot
\int Q_{j}\cdot dx^{j}]=e^{\varphi }\cdot l_{d}^{2}\text{ \ , \ \ \ \ \ \ \
\ \ \ }\widetilde{l}_{d0}^{2}:=l_{d0}^{2}=\text{ const.}
\end{eqnarray*}

For $Q_{j}=-n\cdot \overline{\varphi }_{,j}$, it follows that 
\[
\widetilde{l}_{d}^{2}=\widetilde{l}_{d0}\cdot e^{\varphi -\overline{\varphi }%
}\text{ \ .\ \ } 
\]

If $\widetilde{l}_{d}^{2}$ does not change along the vector field $u$, then $%
\varphi =\overline{\varphi }$ and we have only one dilation field $\varphi
=\varphi (x^{k})=\overline{\varphi }(x^{k})$ which determines the conformal
factor of the metric $\widetilde{g}$, conformal to a Weyl's metric $g$, as
well as the Weyl's covector $Q$. In this case the metric $\widetilde{g}$
appears as a Riemannian metric because of $\widetilde{Q}_{j}=0$. If $\varphi
\neq \overline{\varphi }$, then the metric $\widetilde{g}$, conformal to $g$
is again a Weyl's metric with $\widetilde{Q}_{j}=-n\cdot \widetilde{\varphi }%
_{,j}$ and with $\widetilde{\varphi }=-(\overline{\varphi }-\varphi )$.

{\it Proposition} 2. The necessary and sufficient condition for a metric $%
\widetilde{g}$ conformal ($\widetilde{g}=e^\varphi \cdot g$) to a Weyl's
metric $g$ [obeying the condition $\nabla _\xi g=\frac 1nQ_\xi \cdot g$ for $%
\forall \xi \in T(M)$] to be a Riemannian metric $\widetilde{g}$ [obeying
the condition $\nabla _\xi \widetilde{g}=0$ for $\forall \xi \in T(M)$] is
the condition 
\begin{equation}
Q_\xi =-n\cdot (\xi \varphi )\text{ \ \ \ \ \ , \ \ \ \ }\xi \in T(M)\text{
, \ \ \ \ \ }\varphi \in C^r(M)\text{ , }r\geqq 2\text{ .}  \label{1.29}
\end{equation}

The proof is trivial.

{\it Corollary}. All Riemannian metrics are conformal to a Weyl's metric in
a Weyl's space with $Q_\xi =-n\cdot \xi \varphi $ for $\forall \xi \in T(M)$%
, $\varphi \in C^r(M)$, $r\geqq 2$, [or in a co-ordinate basis with $%
Q_k=-n\cdot $\ $\varphi _{,k}$].

\subsection{Conformal Killing vectors}

A conformal Killing vector field is defined by a recurrent equation.

{\it Definition 2}. A conformal Killing vector field is a contravariant
vector field $u$ obeying the recurrent equation 
\begin{equation}
\pounds _ug=\lambda _u\cdot g\text{ \ \ \ , \ \ \ \ \ \ \ }\lambda _u\in
\otimes _0^0(M)\subset C^r(M)\text{ .}  \label{1.4}
\end{equation}

The equation is called conformal Killing equation. After contracting $%
\pounds _ug$ and $g$ from the last equation with $\overline{g}=g^{ij}\cdot
\partial _i.\partial _j$ [$\partial _i.\partial _j:=\frac 12\cdot (\partial
_i\otimes \partial _j+\partial _j\otimes \partial _i)$] by both basic vector
fields, i.e. after finding the relations 
\begin{equation}
\overline{g}[\pounds _ug]=g^{\overline{k}\overline{l}}\cdot \pounds _ug_{ij}%
\text{ \ \ \ \ \ \ and \ \ \ \ \ \ }\overline{g}[g]=g^{\overline{k}\overline{%
l}}.g_{kl}=n=\dim M\text{ \ ,}  \label{1.5}
\end{equation}
it follows for $\lambda _u$%
\begin{equation}
\lambda _u=\frac 1n\cdot \overline{g}[\pounds _ug]=\frac 2n\cdot \widetilde{f%
}_u\text{ .}  \label{1.6}
\end{equation}

On the other side, the results of the action of the covariant differential
operator $\nabla _u$ and of the Lie differential operator $\pounds _u$ on
the invariant volume element $d\omega $ are recurrent relations for $d\omega 
$ \cite{Manoff-1} 
\begin{equation}
\nabla _u(d\omega )=\overline{f}_u\cdot d\omega \text{ \ , \ \ \ \ }%
\overline{f}_u:=\frac 12\cdot \overline{g}[\nabla _ug]\in \otimes
_0^0(M)\subset C^r(M)\text{ ,}  \label{1.2}
\end{equation}
\begin{equation}
\pounds _u(d\omega )=\widetilde{f}_u\cdot d\omega \text{ \ , \ \ \ \ \ \ \ }%
\widetilde{f}_u:=\frac 12\cdot \overline{g}[\pounds _ug]\in \otimes
_0^0(M)\subset C^r(M)\text{ .}  \label{1.3}
\end{equation}

Therefore, the factor $\lambda _u$ in a conformal Killing vector equation is
related to the factor $\widetilde{f}_u$ by which the invariant volume
element changes when dragged along the same vector field.

\subsection{Problems and results}

On the basis of the proposition $2$ we can state that for {\it every }given
Riemannian metric $\widetilde{g}$ from a Riemannian space and a{\it \ given} 
{\it scalar (dilaton) field} $\varphi (x^k)$ in this space we could generate
a Weyl's metric $g$ in a Weyl's space with the same affine connection as the
affine connection in the Riemannian space. Vice versa, for every given
Weyl's space with Weyl's covariant vector field $Q$ constructed by a scalar
(dilaton) field $\varphi $ and a Weyl's metric $g$ we could generate a
Riemannian metric $\widetilde{g}$ in a Riemannian space with the same affine
connection as in the corresponding Weyl's space.

Therefore, {\it every (metric) tensor-scalar theory of gravitation in a
(pseudo) Riemannian space (with or without torsion) could be reformulated in
a corresponding Weyl's space with Weyl's metric and dilaton field,
generating the Weyl's covector in the Weyl's space} and vice versa: {\it %
every (metric) tensor-scalar theory in a Weyl's space with scalar (dilaton)
field, generating the Weyl's covector, could be reformulated in terms of a
(metric) tensor-scalar theory in the corresponding Riemannian space with the
same affine connections as the affine connection in the Weyl's space.}

The last statement generates the idea of considering the motion of a flow or
the motion of a spinless test particle in a Weyl's space under the following
conditions:

(a) the acceleration of the particle (material element) of the flow is
orthogonal to its velocity,

(b) the velocity of the particle (material element) of the flow is a
non-null shear-free and expansion-free vector

(c) the equation of motion of the particle appears in a special case as an
auto-parallel equation.

The kinematic characteristics of a vector field obeying the condition $%
\nabla _ug-\pounds _ug=0$ can now be considered from a more general point of
view, namely, for spaces with affine connections (which components differ
not only by sign) and then specialized for Weyl's spaces.

\section{Shear-free and expansion-free flows}

Let us now consider conditions for the existence of a shear-free and
expansion-free flow in a $(\overline{L}_n,g)$-space and especially in a
Weyl's space. For this reason we need some facts about

\begin{itemize}
\item  the notion of relative velocity, shear velocity, and expansion
velocity,

\item  conformal Killing vectors,

\item  the properties of Weyl's spaces.
\end{itemize}

\subsection{Deformation velocity, shear velocity, rotation velocity and
expansion velocity}

The notion of relative velocity in a $V_n$-space $(n=4)$ has been introduced
by definition by Ehlers \cite{Ehlers} as the velocity of a particle with
respect to an other particle of its neighborhood in a flow. The physical
interpretation of the notion of relative velocity in $(\overline{L}_n,g)$%
-spaces has been recently discussed in details in \cite{Manoff-preprint}. It
has been shown that the relative velocity $_{rel}v$ is the velocity between
two particles (matter elements) lying at a cross-section (hypersurface)
orthogonal to the velocity $u$ and having equal proper times.

If we introduce the relations

\[
\begin{array}{c}
h_u:=g-\frac 1e\cdot g(u)\otimes g(u)\text{ , \ \ \ \ \ \ \ }h_u=h_{ij}\cdot
e^i.e^j\text{ , \ \ \ \ \ \ \ \ \ \ \ \ \ \ \ }\overline{g}:=g^{ij}\cdot
e_i.e_j, \\ 
\\ 
\nabla _u\xi =\xi ^i\text{ }_{;j}\cdot u^j\cdot e_i\text{ , \thinspace
\thinspace \thinspace \thinspace \thinspace \thinspace \thinspace \thinspace
\thinspace \thinspace \thinspace \thinspace \thinspace \thinspace \thinspace
\thinspace }\xi ^i\text{ }_{;j}:=e_j\xi ^i+\Gamma _{kj}^i\cdot \xi ^k\,\text{%
,\thinspace \thinspace \thinspace \thinspace \thinspace \thinspace
\thinspace \thinspace \thinspace \thinspace \thinspace \thinspace \thinspace
\thinspace \thinspace \thinspace \thinspace \thinspace \thinspace }%
\,\,\,\Gamma _{kj}^i\neq \Gamma _{jk}^i\text{ }, \\ 
\\ 
e:=g(u,u)=g_{\overline{i}\overline{j}}\cdot u^i\cdot u^j=u_{\overline{i}%
}\cdot u^i\neq 0\text{ ,\thinspace \thinspace \thinspace \thinspace
\thinspace \thinspace \thinspace \thinspace \thinspace \thinspace \thinspace
\thinspace \thinspace \thinspace }g(u)=g_{\overline{i}\overline{k}}\cdot
u^k\cdot e^i=u_{\overline{i}}\cdot e^i\text{ ,} \\ 
\\ 
\text{\thinspace }h_{ij}=g_{ij}-\frac 1e\cdot u_i\cdot u_j\text{ ,\thinspace
\thinspace \thinspace \thinspace \thinspace \thinspace \thinspace \thinspace
\thinspace \thinspace \thinspace \thinspace \thinspace }e_i.e_j:=\frac
12\cdot (e_i\otimes e_j+e_j\otimes e_i)\,\,\,\text{,} \\ 
\\ 
h_u(\nabla _u\xi )=h_{i\overline{j}}\cdot \xi ^j\text{ }_{;k}\cdot u^k\cdot
e^i\text{ ,}
\end{array}
\]
the relative velocity $_{rel}v$ \cite{Manoff-preprint}, \cite{Manoff-9}
could be represented in the form 
\begin{eqnarray}
_{rel}v &=&\overline{g}[h_u(\nabla _u\xi )]=g^{ij}\cdot h_{\overline{j}%
\overline{k}}\cdot \xi ^k\,_{;l}\cdot u^l\cdot e_i\text{ \ \ \ ,}
\label{2.11} \\
\,\,\,\,e_i &=&\partial _i\text{ (in a co-ordinate basis),}  \nonumber
\end{eqnarray}
where $\overline{g}[h_u(\xi )]:=\xi _{\perp }=g^{ik}\cdot h_{\overline{k}%
\overline{l}}\cdot \xi ^l\cdot e_i\,$ is called deviation vector field and
(the indices in a co-ordinate and in a non-co-ordinate basis are written in
both cases as Latin indices instead of Latin and Greek indices)

The relative velocity $_{rel}v$ could be written in a $(\overline{L}_n,g)$%
-space under the conditions $g(u,\xi ):=l=0$, $\pounds _\xi u=0$, in the
form \cite{Manoff-9}, \cite{Manoff-9a} 
\[
_{rel}v=\overline{g}[d(\xi )]\text{ .} 
\]

The covariant tensor field $d$ is a generalization for $(\overline{L}_n,g)$%
-spaces of the well known {\it deformation velocity }tensor for $V_n$-spaces 
\cite{Stephani}, \cite{Kramer}. It is usually represented by means of its
three parts: the trace-free symmetric part, called {\it shear velocity }%
tensor (shear), the anti-symmetric part, called {\it rotation velocity }%
tensor (rotation) and the trace part, in which the trace is called {\it %
expansion velocity }(expansion){\it \ }invariant. The physical
interpretation of all parts of the deformation velocity tensor for the
continuous media mechanics in $(\overline{L}_n,g)$-spaces is discussed in 
\cite{Manoff-preprint}.

After some more complicated as for $V_{n}$-spaces calculations, the
deformation velocity tensor $d$ can be given in the form \cite{Manoff-9}

\begin{equation}
d=\sigma +\omega +\frac 1{n-1}\cdot \theta \cdot h_u\text{ .}  \label{2.12}
\end{equation}

The tensor $\sigma $ is the {\it shear velocity} tensor (shear) , 
\begin{equation}
\begin{array}{c}
\sigma =\,_sE-\,_sP=E-P-\frac 1{n-1}\cdot \overline{g}[E-P]\cdot h_u=\sigma
_{ij}\cdot e^i.e^j= \\ 
\\ 
=E-P-\frac 1{n-1}\cdot (\theta _o-\theta _1)\cdot h_u\text{ ,}
\end{array}
\end{equation}

\noindent where

\begin{equation}
\begin{array}{c}
_sE=E-\frac 1{n-1}\cdot \overline{g}[E]\cdot h_u\text{ , \thinspace
\thinspace \thinspace \thinspace \thinspace \thinspace \thinspace \thinspace
\thinspace \thinspace \thinspace \thinspace \thinspace \thinspace }\overline{%
g}[E]=g^{\overline{i}\overline{j}}\cdot E_{ij}=\theta _o\text{ ,} \\ 
\\ 
E=h_u(\varepsilon )h_u\text{ , \thinspace \thinspace \thinspace \thinspace
\thinspace \thinspace \thinspace \thinspace \thinspace \thinspace \thinspace
\thinspace }k_s=\varepsilon -m\text{ , \thinspace \thinspace \thinspace
\thinspace \thinspace \thinspace \thinspace \thinspace \thinspace \thinspace
\thinspace }\varepsilon =\frac 12(u_{\text{ };l}^i\cdot g^{lj}+u_{\text{ }%
;l}^j\cdot g^{li})\cdot e_i.e_j\text{ ,} \\ 
\\ 
m=\frac 12(T_{lk}\,^i\cdot u^k\cdot g^{lj}+T_{lk}\,^j\cdot u^k\cdot
g^{li})\cdot e_i.e_j\text{ .}
\end{array}
\label{2.14}
\end{equation}

The tensor $_sE$ is the {\it torsion-free shear velocity} {\it \ }tensor,
the tensor $_sP$ is the {\it shear velocity} tensor {\it induced by the
torsion},

{\it 
\begin{equation}
\begin{array}{c}
_sP=P-\frac 1{n-1}\cdot \overline{g}[P]\cdot h_u\text{ , \thinspace
\thinspace \thinspace \thinspace \thinspace \thinspace \thinspace \thinspace
\thinspace \thinspace \thinspace \thinspace \thinspace \thinspace \thinspace 
}\overline{g}[P]=g^{\overline{k}\overline{l}}\cdot P_{kl}=\theta _1\text{,
\thinspace \thinspace \thinspace \thinspace \thinspace \thinspace }%
P=h_u(m)h_u\text{ ,} \\ 
\\ 
\theta _1=T_{kl}\,^k\cdot u^l\text{ ,\thinspace \thinspace \thinspace
\thinspace \thinspace \thinspace \thinspace }\theta _o=u^n\text{ }%
_{;n}-\frac 1{2e}(e_{,k}\cdot u^k-g_{kl;m}\cdot u^m\cdot u^{\overline{k}%
}\cdot u^{\overline{l}})\text{ ,\thinspace \thinspace \thinspace }\theta
=\theta _o-\theta _1\text{ . }
\end{array}
\label{2.15}
\end{equation}
}

The invariant $\theta $ is the {\it expansion velocity,} the invariant{\it \ 
$\theta _o$} is the {\it torsion-free expansion velocity,} the invariant $%
\theta _1$ is the {\it expansion velocity induced by the torsion.}

The tensor $\omega $ is the {\it \ rotation velocity }tensor (rotation
velocity),

\begin{equation}
\begin{array}{c}
\omega =h_u(k_a)h_u=h_u(s)h_u-h_u(q)h_u=S-Q\text{ ,} \\ 
\\ 
s=\frac 12(u^k\text{ }_{;m}\cdot g^{ml}-u^l\text{ }_{;m}\cdot g^{mk})\cdot
e_k\wedge e_l\text{ ,} \\ 
\\ 
q=\frac 12(T_{mn}\,^k\cdot g^{ml}-T_{mn}\,^l\cdot g^{mk})\cdot u^n\cdot
e_k\wedge e_l\text{ , \ } \\ 
\\ 
\text{\thinspace \thinspace }S=h_u(s)h_u\text{ , \thinspace \thinspace
\thinspace \thinspace \thinspace \thinspace \thinspace \thinspace \thinspace
\thinspace \thinspace \thinspace \thinspace \thinspace \thinspace \thinspace
\thinspace \thinspace \thinspace }Q=h_u(q)h_u\text{ .}
\end{array}
\label{2.16}
\end{equation}

The tensor $S$ is the {\it torsion-free rotation velocity} tensor, the
tensor $Q$ is the {\it \ rotation velocity }tensor {\it induced by the
torsion.}

By means of the expressions for $\sigma $, $\omega $ and $\theta $ the
deformation velocity tensor $d$ can be decomposed in two parts: $d_0$ and $%
d_1$

\begin{equation}
d=d_{o}-d_{1}\text{ , \ \ \ \thinspace \thinspace \thinspace }%
d_{o}=\,_{s}E+S+\frac{1}{n-1}.\theta _{o}.h_{u}\text{ , \ \ \thinspace
\thinspace \thinspace \thinspace }d_{1}=\,_{s}P+Q+\frac{1}{n-1}.\theta
_{1}.h_{u}\text{ ,}  \label{2.17}
\end{equation}

\noindent where $d_o$ is the {\it torsion-free deformation velocity} tensor
and $d_1$ is the {\it deformation velocity }tensor {\it induced by the
torsion. }For the case of $V_n$-spaces $d_1=0$ ($_sP=0$, $Q=0$, $\theta _1=0$%
).

After some calculations, the shear velocity tensor $\sigma $ and the
expansion velocity $\theta $ can also be written in the forms

\begin{equation}
\begin{array}{c}
\sigma =\frac{1}{2}\{h_{u}(\nabla _{u}\overline{g}-\pounds _{u}\overline{g}%
)h_{u}-\frac{1}{n-1}\cdot (h_{u}[\nabla _{u}\overline{g}-\pounds _{u}%
\overline{g}])\cdot h_{u}\}\text{ }= \\ 
\\ 
=\frac{1}{2}\{h_{i\overline{k}}\cdot (g^{kl}\text{ }_{;m}\cdot u^{m}-\pounds
_{u}g^{kl})\cdot h_{\overline{l}j}-\frac{1}{n-1}\cdot h_{\overline{k}%
\overline{l}}\cdot (g^{kl}\text{ }_{;m}\cdot u^{m}-\pounds _{u}g^{kl})\cdot
h_{ij}\}\cdot e^{i}.e^{j}\text{ ,}
\end{array}
\label{2.18}
\end{equation}
\begin{equation}
\theta =\frac{1}{2}\cdot h_{u}[\nabla _{u}\overline{g}-\pounds _{u}\overline{%
g}]=\frac{1}{2}h_{\overline{i}\overline{j}}\cdot (g^{ij}\text{ }_{;k}\cdot
u^{k}-\pounds _{u}g^{ij})\text{ .}  \label{2.19}
\end{equation}

The physical interpretation of the velocity tensors $d$, $\sigma $, $\omega $%
, and of the invariant $\theta $ for the case of $V_4$-spaces \cite{Synge}, 
\index{Synge J. L.@Synge J. L.}\cite{Ehlers}%
\index{Ehlers J.@Ehlers J.}, can also be extended for $(%
\overline{L}_4,g)$-spaces (for more details see \cite{Manoff-preprint}). It
is easy to be seen that the existence of some kinematic characteristics ($%
_sP $, $Q$, $\theta _1$) depends on the existence of the torsion tensor
field. They vanish if it is equal to zero (e.g. in $V_n$-spaces).

{\it Remark. }It should be stressed that the decomposition of the
deformation tensor $d$ could not follow from the decomposition of $%
u^i\,_{;j} $ as this has been done by Ehlers \cite{Ehlers} for (pseudo)
Riemannian spaces without torsion ($V_n$-spaces, $n=4$). In $(\overline{L}%
_n,g)$-spaces 
\begin{eqnarray}
u^i\,_{;j} &=&\frac 1e\cdot a^i\cdot u_{\overline{j}}+g^{ik}\cdot (_sE_{%
\overline{k}\overline{j}}+S_{\overline{k}\overline{j}}+\frac 1{n-1}\cdot
\theta _0\cdot h_{\overline{k}\overline{j}})+  \nonumber \\
&&  \nonumber \\
&&+\frac 1{2\cdot e}\cdot u^i\cdot (e_{,l}-g_{mn;l}\cdot u^{\overline{m}%
}\cdot u^{\overline{n}})\cdot h^{lk}\cdot g_{\overline{k}\overline{j}}\text{
\ .}  \label{2.19a}
\end{eqnarray}

We can now introduce the notion of shear-free and expansion-free flow.

{\it Definition 3.} A flow is called a shear-free and expansion-free flow if
it has no shear velocity $\sigma $ and no expansion velocity $\theta $, i.e.
a flow is a shear-free and expansion-free flow if $\sigma =0$ and $\theta =0$%
.

The representation of the shear velocity tensor $\sigma $ and the expansion
invariant $\theta $ by means of the covariant and Lie derivatives of the
contravariant metric tensor $\overline{g}$ give rise to some important
conclusions about their vanishing or nonvanishing in a $(\overline{L}_n,g)$%
-space.

From the structure of $\sigma $ and $\theta $ in the expressions (\ref{2.18}%
) and (\ref{2.19}) respectively, it is obviously that if $\pounds _u%
\overline{g}=\nabla _u\overline{g}$ then $\sigma =0$ and $\theta =0$, i.e.
the condition $\pounds _u\overline{g}=\nabla _u\overline{g}$ appears as a
sufficient conditions for $\sigma =0$ and $\theta =0$. On the other side,
this condition could be written in the form $\pounds _ug=\nabla _ug$ because
of the relations $\pounds _u\overline{g}=-\overline{g}(\pounds _ug)\overline{%
g}$ and $\nabla _u\overline{g}=-\overline{g}(\nabla _ug)\overline{g}$.

On the basis of the above considerations we could now formulate the
following proposition:

{\it Proposition 3}. If a metric $g$ in a space with affine connections and
metrics [a $(\overline{L}_n,g)$- or a $(L_n,g)$-space] fulfills the
condition 
\begin{equation}
\pounds _u\overline{g}=\nabla _u\overline{g}\text{ \ \ or \ \ \ \ \ \ }%
\pounds _ug=\nabla _ug\text{ , }
\end{equation}
then the space admits a non-null shear-free and expansion-free contravariant
vector field $u$.

\subsection{Shear-free and expansion-free velocity vector}

If we now compare the recurrent relations obtained as a result of the action
of the covariant differential operator $\nabla _u$ and of the Lie
differential operator $\pounds _u$ on the invariant volume element $d\omega $
the question could arise what are the conditions for the equivalence of the
action of both the differential operators on $d\omega $, i.e. under which
conditions for the vector field $u$ and the metric $g$ we could have the
relation 
\begin{equation}
\nabla _u(d\omega )=\pounds _u(d\omega )\text{ ,}  \label{1.12}
\end{equation}
\noindent which is equivalent to the relation 
\begin{equation}
\overline{g}[\nabla _ug]=\overline{g}[\pounds _ug]\text{ \ \ \ \ or \ \ \ \ }%
\overline{g}[\nabla _ug-\pounds _ug]=0\text{ \ .}  \label{1.13}
\end{equation}

What does this relation physically mean? The action of the covariant
differential operator $\nabla _u$ is determined only on a curve at which $u$
is a tangent vector. The dragging of a contravariant vector $\xi $ along $u$
requires the existence of $\nabla _\xi u$ as the change of the vector $u$ by
a transport along $\xi $ [because of the relation $\pounds _u\xi =\nabla
_u\xi -\nabla _\xi u-T(u,\xi )$]. The existence of $\nabla _\xi u$ is
related to the condition that the components $\xi ^i(x^k)$ of the vector $%
\xi =\xi ^i\cdot \partial _i$ should be differentiable functions on
neighborhoods of the points of the curve with tangent vector $u$. This
condition is not necessary for a transport of $\xi $ along $u$. On this
ground, the dragging of $\xi $ along $u$ determines neighborhoods of the
points of the curve with tangent vector $u$. These neighborhoods are moving
with the flow along its velocity vector $u$.

The transport of $g$ is only on the curve and not on the vicinities out of
the points of the curve. The action of the Lie differential operator is
determined on the vicinities on and out of the points of the curve with
tangent vector $u$ on it. The dragging-along of $g$ is on the whole
vicinities of the points of the curve and not only along the curve. If the
dragging-along $u$ of $g$ is equal to the transport of $g$ along $u$, then
an observer with its worldline as the curve with tangent vector $u$ could
not observe any changes in its worldline vicinity different from those who
could register on its worldline. The observer will see its surroundings as
if they are moving with him.

It is obvious that in the general case, in $(\overline{L}_n,g)$-spaces, a
sufficient condition for fulfilling the last relation is the condition 
\begin{equation}
\nabla _ug-\pounds _ug=0\text{ \ .}  \label{1.14}
\end{equation}

Let us now discuss in brief the physical interpretation of the last
condition.

The change of the length of a vector field $u$ along the curve to which it
is a tangent vector field could be written for $(\overline{L}_n,g)$-spaces
in the form \cite{Manoff-e} 
\[
ul_u=\pm \frac 1{2\cdot l_u}\cdot [(\nabla _ug)(u,u)+2\cdot g(\nabla _uu,u)]%
\text{ \thinspace \thinspace \thinspace \thinspace ,\thinspace \thinspace
\thinspace \thinspace \thinspace \thinspace \thinspace \thinspace \thinspace
\thinspace \thinspace }l_u\neq 0\,\,\text{,} 
\]

\noindent or in the form 
\[
ul_u=\pm \frac 1{2\cdot l_u}\cdot (\pounds _ug)(u,u)\,\,\,\text{.} 
\]

Therefore, 
\[
(\nabla _ug)(u,u)+2\cdot g(a,u)=(\pounds _ug)(u,u)\,\,\,\,\,\,\text{%
,\thinspace \thinspace \thinspace \thinspace \thinspace \thinspace
\thinspace \thinspace \thinspace \thinspace \thinspace \thinspace \thinspace 
}a=\nabla _uu\text{ \thinspace \thinspace \thinspace .} 
\]

If the relation $\nabla _ug=\pounds _ug$ is fulfilled then it follows from
the last relation that $g(a,u)=0$.\ If the vector field $u$ is interpreted
as the velocity of a particle (matter element) then the vector field $a$ is
interpreted as its acceleration. Therefore, the condition $\nabla
_ug=\pounds _ug$ is a sufficient condition for the acceleration $a$ of a
particle to be orthogonal to its velocity $u$.

{\it Remark}. In $U_n$- and $V_n$-spaces this condition is automatically
fulfilled if $u$ is considered as a normalized vector field $[g(u,u)=e=e_0=$
const.$\neq 0]$. In more sophisticated spaces this condition is not
fulfilled identically even if $u$ is a normalized vector field because of
the relation (see the above expression for $ul_u$) 
\[
g(a,u)=\frac 12\cdot [ue-(\nabla _ug)(u,u)]\text{ .} 
\]

{\it Remark}. The last expression could be interpreted physically as a
criteria for the non-orthogonality of the acceleration $a$ and the velocity $%
u$ in a $(\overline{L}_n,g)$-space. If $e=g(u,u):=$ const. then $ue=0$ and
the non-metricity $(\nabla _ug\neq 0)$ could be considered as a direct
criteria for the non-orthogonality of $a$ and $u$%
\[
g(a,u)=\frac 12\cdot (\nabla _ug)(u,u)\,\,\,\text{.} 
\]

Therefore, the imposition of the condition $\nabla _ug=\pounds _ug$
restricts the possible motions of a particle with velocity $u$ in a $(%
\overline{L}_n,g)$-space to motions with acceleration $a$ orthogonal to the
velocity $u$.

On the other side, the Killing equation is usually related to symmetries of
the metric tensor $g$. If a Killing vector exists then the metric, written
in appropriate co-ordinates so that the Killing vector is tangent to one of
them, will not depend on this co-ordinate. From physical point of view the
Killing equation $\pounds _ug=0$ for the vector $u$ leads to conservation of
the length $l_u$ of $u$ along the same vector, i.e. $\pounds _ug=0$ is a
sufficient condition for preservation of the velocity vector $u$ along the
trajectory of a particle moving with velocity $u$.

The condition (\ref{1.14}) is fulfill:

(a) in Riemannian spaces (with or without torsion) [for which $\nabla
_{u}g=0 $ for $\forall u\in T(M)$], when the Killing equation \cite{Yano} 
\begin{equation}
\pounds _{u}g=0\text{ \ }  \label{1.15}
\end{equation}
is fulfilled for the vector field $u$.

(b) in Weyl's spaces (with or without torsion) [for which $\nabla _{u}g=%
\frac{1}{n}\cdot Q_{u}\cdot g$ for $\forall u\in T(M)$], when the conformal
Killing equation 
\begin{equation}
\pounds _{u}g=\lambda _{u}\cdot g\text{ \ \ \ \ \ \ with \ \ \ \ \ \ }%
\lambda _{u}=\frac{1}{n}\cdot Q_{u}  \label{1.16}
\end{equation}
is fulfilled for the vector field $u$.

In $(\overline{L}_n,g)$-spaces, the relation $\nabla _ug-\pounds _ug=0$
could be written in a co-ordinate basis in the form 
\begin{eqnarray}
\pounds _ug_{ij} &=&g_{ij;k}\cdot u^k+g_{kj}\cdot u^{\overline{k}}\,_{;%
\underline{i}}+g_{ik}\cdot u^{\overline{k}}\,_{;\underline{j}}+(g_{kj}\cdot
T_{l\underline{i}}\,^{\overline{k}}+g_{ik}\cdot T_{l\underline{j}}\,^{%
\overline{k}})\cdot u^l=  \nonumber \\
&& \\
&=&g_{ij;k}\cdot u^k\text{ ,}  \label{1.17}
\end{eqnarray}
or in the forms 
\begin{equation}
g_{kj}\cdot u^{\overline{k}}\,_{;\underline{i}}+g_{ik}\cdot u^{\overline{k}%
}\,_{;\underline{j}}+(g_{kj}\cdot T_{l\underline{i}}\,^{\overline{k}%
}+g_{ik}\cdot T_{l\underline{j}}\,^{\overline{k}})\cdot u^l=0\text{ \ ,}
\label{1.18}
\end{equation}
\begin{equation}
g_{kj}\cdot (u^{\overline{k}}\,_{;\underline{i}}-T_{\underline{i}l}\,^{%
\overline{k}}\cdot u^l)+g_{ik}\cdot (u^{\overline{k}}\,_{;\underline{j}}-T_{%
\underline{j}l}\,^{\overline{k}}\cdot u^l)=0\text{ ,}  \label{1.19}
\end{equation}
where 
\[
T_{\underline{i}l}\,^{\overline{k}}:=f_i\,^m\cdot T_{ml}\,^n\cdot f^k\,_n%
\text{, \ \ \ \ \ \ \ \ \ \ \ \ \ \ }\ \ u^{\overline{k}}\,_{;\underline{j}%
}:=f^k\,_l\cdot u^l\,_{;m}\cdot f_j\,^m. 
\]

The equation (\ref{1.19}) could be called ''generalized conformal Killing
equation'' for the vector field $u$ in the case $\pounds _{u}g=\nabla _{u}g$.

After multiplication of the last expression, equivalent to $\pounds
_{u}g_{ij}=g_{ij;k}\cdot u^{k}$, with $u^{\overline{i}}$ and $g^{m\overline{j%
}}$ and summation over $\overline{i}$ and $\overline{j}$ (and then changing
the index $m$ with $i$) we can find the equation for the vector field $u$ in
a co-ordinate basis 
\begin{equation}
u^{i}\,_{;j}\cdot u^{j}+g_{\overline{l}\overline{k}}\cdot u^{l}\cdot
(u^{k}\,_{;j}-T_{jm}\,^{k}\cdot u^{m})\cdot g^{ji}=0\text{ \ ,}  \label{1.20}
\end{equation}
or as an equation for the acceleration $a=a^{i}\cdot \partial _{i}$ with 
\begin{equation}
a^{i}=-g_{\overline{l}\overline{k}}\cdot u^{l}\cdot
(u^{k}\,_{;j}-T_{jm}\,^{k}\cdot u^{m})\cdot g^{ji}=-g_{\overline{l}\overline{%
k}}\cdot u^{l}\cdot k^{ki}\text{ \ ,}  \label{1.21}
\end{equation}
where 
\begin{equation}
k^{ki}=(u^{k}\,_{;j}-T_{jm}\,^{k}\cdot u^{m})\cdot g^{ji}\text{ .}
\label{1.22}
\end{equation}

On the other side, from the equation (\ref{1.19}) it is obvious that a
sufficient condition for fulfilling the equation (\ref{1.19}) is the
condition for $u^{i}$%
\begin{equation}
u^{k}\,_{;j}-T_{jl}\,^{k}\cdot u^{l}=0\text{ \ \ \ \ \ \ \ or \ \ \ \ \ }%
u^{k}\,_{;j}=T_{jl}\,^{k}\cdot u^{l}\text{ \ \ .\ }  \label{1.23}
\end{equation}

From the last expression, it follows that if the vector $u$ fulfills this
condition it should be an auto-parallel vector field since 
\begin{equation}
u^{i}\,_{;j}\cdot u^{j}=a^{i}=0\text{ \ \ \ .}  \label{1.24}
\end{equation}

If (\ref{1.23}) \ is fulfilled, then the following relations are valid: 
\begin{eqnarray*}
u^m\cdot R_{\,\,\;mkl}^i
&=&-(T_{ml}\,^i\,_{;k}-T_{mk}\,^i\,_{;l}+T_{ml}\,^n\cdot T_{nk}\,^i- \\
&& \\
&&-T_{mk}\,^n\cdot T_{nl}\,^i+T_{kl}\,^n\cdot T_{mn}\,^i)\cdot u^m\text{ \ ,}
\end{eqnarray*}
\[
R_{mk}\cdot u^m\cdot u^k=g_i^l\cdot R^i\,_{mkl}\cdot u^m\cdot
u^k=T_{lm}\,^l\,_{;k}\cdot u^k\cdot u^m\text{ \ \ .} 
\]

On the basis of the above relations, we can prove the preposition:

{\it Proposition 4}. If a contravariant non-null vector field $u$ in a space
with affine connections and metrics [a $(\overline{L}_n,g)$- or a $(L_n,g)$%
-space] fulfills the equation (\ref{1.23}) 
\[
u^k\,_{;j}-T_{jl}\,^k\cdot u^l=0\text{ \ \ \ \ \ \ \ or \ \ \ \ \ }%
u^k\,_{;j}=T_{jl}\,^k\cdot u^l\text{ ,} 
\]
then this vector field is an auto-parallel shear-free and expansion-free
vector field.

Let us now consider a Weyl's space admitting the condition (\ref{1.14}).

For Weyl's spaces ($\nabla _ug=\frac 1n\cdot Q_u\cdot g$) the condition (\ref
{1.14}) degenerate in the condition for the existence of a conformal Killing
vector $u$%
\begin{equation}
\pounds _ug=\lambda _u\cdot g\text{ \ \ \ \ \ \ with \ \ \ \ }\lambda
_u=\frac 1n\cdot Q_u\text{ .}  \label{2.20}
\end{equation}

Therefore, we can prove the following propositions:

{\it Proposition 5}. If a contravariant non-null vector field in a Weyl's
space fulfills a conformal Killing equation of the type 
\begin{equation}
\pounds _ug=\lambda _u\cdot g\text{ \ \ \ \ \ \ with \ \ \ \ }\lambda
_u=\frac 1n\cdot Q_u\text{ ,}  \label{2.22}
\end{equation}
then this conformal vector field $u$ is also a shear-free and expansion-free
vector field.

{\it Proposition 6}. If a contravariant non-null vector field $u$ in a
Weyl's space fulfills the equation (\ref{1.23}) 
\[
u^{k}\,_{;j}-T_{jl}\,^{k}\cdot u^{l}=0\text{ \ \ ,} 
\]
then it is an auto-parallel, shear-free and expansion-free conformal Killing
vector field.

The auto-parallel equation (\ref{1.24}) for the vector field $u$ is
interpreted as an equation of motion for a free spinless test particle in
spaces with affine connection and metrics \cite{Manoff-6}, \cite{Manoff-7}.
Let us now consider this equation more closely in Weyl's spaces.

\section{Auto-parallel equation in Weyl's spaces as an equation for a free
moving spinless test particles}

Every (covariant) gravitational theory in spaces with affine connections and
metrics should obey a condition (equation) for description of the motion of
a free moving spinless test particle.

The notion of a (spinless) test particle is usually related to a material
point moving in an external field or in space-time without changing the
characteristics of the external field or of the space-time. This means that
the dynamical characteristics of the test particle does not generate
additional influence on the particle's motion under external conditions. If
a (spinless) test particle is moving in the space-time it does not act on
its geometric properties related in some gravitational theories to the
properties of other material points, distributions and dynamical systems on
the basis of field equations. These field equations describe the evolution
of the physical structures in the space-time and vice versa: the evolution
of the space-time under the existence of physical systems in it. Under these
conditions, it is assumed that the motion of a (spinless) test particle
depends explicitly only on the geometric characteristics of the mathematical
model describing the space-time and only implicitly on the field equations
determining the geometric characteristics of the space-time. On this basis,
we can consider all characteristics of a space-time model as given and
ignore the structure of a concrete field theory (field equations)
determining them.

In (pseudo) Riemannian spaces without torsion ($V_n$-spaces) $(n=4)$ the
motion of a free spinless particle is described by the use of a geodesic
line identical in this type of spaces with an auto-parallel trajectory. It
is {\it a belief} of some authors that a geodesic line should also be the
trajectory of a free moving spinless test particle in $(L_n,g)$- and $(%
\overline{L}_n,g)$-spaces where geodesics are different from auto-parallel
trajectories. The reasons for this belief are the different constructions of
theories where the difference between geodesics and auto-parallels has been
seen as related to the existence of forces generated by the space-time.
Recently, it has been proved \cite{Manoff-7} that a free spinless test
particle could move on an auto-parallel trajectory in a $(L_n,g)$- or in a $(%
\overline{L}_n,g)$-spaces. This fact is based on the principle of
equivalence \cite{Iliev-1}$\div $\cite{Hartley} and could not be ignored in
physical investigations \cite{Manoff-frame}. Moreover, the presence of an
additional force term (defined by non-metricity) does not mean that a
particle has an additional charge (see, for instance, inertial forces in $%
V_n $- and $E_n$-spaces) because this term could be removed by the use of an
appropriate extended differential operator generating a new affine
connection and a new frame of reference in space-time \cite{Manoff-frame}.

Usually the following definition of a free moving test particle in a space
with affine connections and metrics [and especially in (pseudo) Riemannian
spaces without torsion] is introduced \cite{Borisova-1}:

{\it Definition 4}. A free spinless test particle is a material point with
rest mass (density) $\rho $, velocity $u$ (as tangent vector $u$ to its
trajectory) and momentum (density) $p:=\rho \cdot u$ with the following
properties:

(a) The momentum density $p$ does not change its direction along the world
line of the material point, i.e. the vector $p$ fulfills the recurrent
condition $\nabla _{u}p=f\cdot p$, or the condition $\nabla _{u}p=0$, as
conditions for parallel transport along $u$.

(b) The momentum density $p$ does not change its length $l_{p}=\mid
g(p,p)\mid ^{1/2}$ along the world line of the material point, i.e. $p$
fulfills the condition $ul_{p}=0$.

The change of the length of a vector field $p$ along a vector field $u$ in a 
$(\overline{L}_n,g)$-space could be found in the form \cite{Manoff-e} 
\begin{equation}
ul_p=\pm \frac 1{2\cdot l_p}\cdot [(\nabla _ug)(p,p)+2\cdot g(\nabla _up,p)]%
\text{ \ \ , \ \ \ \ \ \ \ \ \ }l_p\neq 0\text{ \ .}  \label{4.1}
\end{equation}

{\it Remark}. The basis for this definition is the consideration of the
notion of momentum density in a classical field theory in spaces with affine
connections and metrics \cite{Manoff-8}, \cite{Manoff-8b}. A material point
(mass element) is characterized by its rest mass density $\rho $ and
velocity $u$. A material point (particle) which does not interact with its
surroundings should have energy-momentum tensor $G$ of the type $(G)%
\overline{g}=u\otimes p$ with $p=\rho \cdot u$ and $p$ should not change
along its trajectory with tangent vector $u$.

Let us now consider the two conditions (a) and (b) for $p$ separately to
each other.

(a) If we write $p$ in its explicit form $p=\rho \cdot u$, then the
condition for a parallel transport of $p$ along $u$ could be written as 
\begin{equation}
\nabla _{u}u=[f-u(\log \rho )]\cdot u\text{ ,}  \label{4.2}
\end{equation}
with 
\begin{equation}
f=u(\log \rho )+\frac{1}{2\cdot e}\cdot \lbrack ue-(\nabla _{u}g)(u,u)]\text{
\ , \ \ \ \ \ \ }e=g(u,u)\neq 0\text{ ,}  \label{4.3}
\end{equation}
and 
\begin{equation}
\nabla _{u}u=\frac{1}{2\cdot e}\cdot \lbrack ue-(\nabla _{u}g)(u,u)]\cdot u%
\text{ .}  \tag{(a)}  \label{4.4}
\end{equation}

In the special case of \ (pseudo) Riemannian spaces ($V_{n}$- or $U_{n}$-
spaces), where $\nabla _{u}g=0$, $e=$ const. $\neq 0$, $\rho =\,$const., $%
f=0 $, it follows that $\nabla _{u}p=0$, and $\nabla _{u}u=0$. At the same
time, $ul_{p}=0$. The parallel equation $\nabla _{u}p=0$ has as a corollary
the preservation of the length $l_{p}$ of $p$ along $u$. This is not the
case if a space is not a (pseudo) Riemannian space.

(b) The conservation of the momentum density $p$ along the trajectory of the
particles $[ul_{p}=0]$ requires the transport of $p$ on this trajectory to
be of the type of a Fermi-Walker transport, i.e. $p$ should obey an equation
of the type \cite{Manoff-2}, \cite{Manoff-3} 
\begin{equation}
\nabla _{u}p=\overline{g}(^{F}\omega -\frac{1}{2}\cdot \nabla _{u}g)(p)=%
\overline{g}[^{F}\omega (p)]-\frac{1}{2}\cdot \overline{g}(\nabla _{u}g)(p)%
\text{ ,}  \label{3.3}
\end{equation}
where $^{F}\omega \in \Lambda ^{2}(M)$ is an antisymmetric tensor of 2nd
rank. For a free particle it could be related to the rotation velocity
tensor (\ref{2.16}) of the velocity $u$, i.e. $^{F}\omega :=\omega $. Then $%
\omega (p)=0$ [because of $\omega (\rho \cdot u)=\rho \cdot (\omega (u))=0$]
and we have for $\nabla _{u}p$%
\begin{equation}
\nabla _{u}p=-\frac{1}{2}\cdot \overline{g}(\nabla _{u}g)(p)\text{ , \ \ \ \
\ \ \ \ }ul_{p}=0\text{ \ .}  \label{3.4}
\end{equation}

\bigskip For the vector field $u$ follows the corresponding condition

\begin{equation}
\nabla _{u}u=-\{[u(\log \rho )]\cdot u+\frac{1}{2}\cdot \overline{g}(\nabla
_{u}g)(u)\}\text{ \ .}  \tag{(b)}  \label{3.5}
\end{equation}

Therefore, \ for $u$ we have two equations as corollaries from the
requirements for the momentum density $p$: equation (a) which follows from
the condition for preservation of the direction of the momentum density $p$,
and equation (b) which follows from the condition for preservation of the
length $l_{p}$ of the momentum density $p$.

(a) The first equation (a) \ for $u$ \ is the auto-parallel equation in its
non-canonical form. It does not depend on the rest mass density $\rho $ of
the particle. From the equation, it follows that the necessary and
sufficient condition for a spinless test particle to move in a space with
affine connections and metrics on a trajectory described by the
auto-parallel equation in its canonical form $(\nabla _uu=0)$ is the
condition 
\begin{equation}
\lbrack ue-(\nabla _ug)(u,u)]\cdot u=0\text{ \ \ \ , \ \ \ \ \ \ }e\neq 0%
\text{ .}  \label{4.7}
\end{equation}

Since $g(u,u)=e\neq 0$, after contracting the equation with $g(u)$, we
obtain the condition 
\begin{equation}
ue-(\nabla _{u}g)(u,u)=0\text{ , \ \ \ \ or \ \ \ \ \ }ue=(\nabla _{u}g)(u,u)%
\text{ \ \ \ .}  \label{4.8}
\end{equation}

This condition determines how the length of the vector $u$ should change
with respect to the change of the metric $g$ along $u$ if $u$ should be an
auto-parallel vector field with $\nabla _{u}u=0$.

If we consider a Weyl's space as a model of space-time, this condition will
take the form 
\begin{equation}
ue=\frac{1}{n}\cdot Q_{u}\cdot e\text{ \ \ , or \ \ \ \ \ \ }u(\log e)=\frac{%
1}{n}\cdot Q_{u}\text{ \ \ ,\ \ }  \label{4.9}
\end{equation}
leading to the relation for $e$%
\begin{equation}
e=e_{0}\cdot \exp (\frac{1}{n}\cdot \int Q_{u}\cdot d\tau )\text{ \ \ , \ \
\ \ \ \ \ \ \ \ \ \ \ }e_{0}=\,\text{const.,}  \label{4.10}
\end{equation}
where $u=d/d\tau $ and $\tau =\tau (x^{k})$ is the canonical parameter of
the trajectory of the particle.

If $Q_{u}$ is constructed by the use of a dilaton field $\overline{\varphi }$
as $Q_{u}=-d\overline{\varphi }/d\tau $, then $e$ would change under the
condition 
\begin{equation}
e=e_{0}\cdot \exp (-\frac{1}{n}\cdot \overline{\varphi })\text{ .}
\label{4.11}
\end{equation}

The dilaton field $\overline{\varphi }$ could be represented by means of $e$
in the form 
\begin{equation}
\overline{\varphi }=-n\cdot \log (\frac{e}{e_{0}})\text{ .}  \label{4.12}
\end{equation}

Therefore, the dilaton field $\overline{\varphi }$ takes the role of a
length scaling factor for the velocity of a test particle.

(b) From the second equation (b) for $u$, it follows that a necessary and
sufficient condition for a free spinless test particle to move in a space
with affine connections and metrics on a trajectory, described by the
auto-parallel equation in its canonical form ($\nabla _{u}u=0$), is the
condition 
\begin{equation}
\lbrack u(\log \rho )]\cdot u=-\frac{1}{2}\cdot \overline{g}(\nabla _{u}g)(u)%
\text{ \ \ .}  \label{3.6}
\end{equation}

Since $g(u,u)=e\neq 0$, \ after contracting the last equation with $g(u)$ we
obtain the condition 
\[
u(\log \rho )\cdot e=-\frac{1}{2}\cdot \overline{g}(\nabla _{u}g)(u)[g(u)]=-%
\frac{1}{2}\cdot (\nabla _{u}g)(u,u)\text{\ \ ,} 
\]
or 
\begin{equation}
u(\log \rho )=-\frac{1}{2\cdot e}\cdot (\nabla _{u}g)(u,u)\text{ .}
\label{3.6a}
\end{equation}

For $u=d/d\tau $, it follows the equation 
\[
\frac{d}{d\tau }(\log \rho )=-\frac{1}{2\cdot e}\cdot (\nabla _{u}g)(u,u)%
\text{ ,} 
\]
with the solution for $\rho (x^{k}(\tau ))$%
\[
\rho =\rho _{0}\cdot \exp (-\frac{1}{2}\cdot \int \frac{1}{e}\cdot (\nabla
_{u}g)(u,u)\cdot d\tau )\text{ \ .} 
\]

The last condition is for the rest mass density $\rho $ which it has to obey
if the particle should move on an auto-parallel trajectory or if we observe
the motion of a particle as a free motion in the corresponding space
considered as a mathematical model of the space-time.

If we consider a Weyl's space as a model of the space-time, the condition (%
\ref{3.6}) will take the form 
\begin{equation}
\lbrack u(\log \rho )]\cdot u=-\frac{1}{2\cdot n}\cdot Q_{u}\cdot u\text{ ,}
\label{3.7}
\end{equation}
or the form 
\[
\lbrack u(\log \rho )+\frac{1}{2\cdot n}\cdot Q_{u}]\cdot u=0\text{ .} 
\]

Since $g(u,u)=e\neq 0$, \ after contracting the last equation with $g(u)$ we
obtain the condition 
\begin{equation}
u(\log \rho )+\frac{1}{2\cdot n}\cdot Q_{u}=0\text{ \ .}  \label{3.8}
\end{equation}

Therefore, the rest mass density $\rho $ should change on the auto-parallel
trajectory of the particle as 
\begin{equation}
\rho =\rho _0\cdot \exp \left[ -\frac 1{2\cdot n}\cdot \int Q_u\cdot d\tau %
\right] \text{ , \ \ \ \ \ \ \ \ }\rho _0=\text{const. }  \label{3.9}
\end{equation}

Furthermore, if $Q_{u}$ is constructed by the use of a dilaton field $%
\overline{\varphi }$ as $Q_{u}=-d\overline{\varphi }/d\tau $, then $\rho $
would change under the condition 
\begin{equation}
\rho =\rho _{0}\cdot \exp \left[ \frac{1}{2\cdot n}\cdot \int \frac{d%
\overline{\varphi }}{d\tau }\cdot d\tau \right] =\rho =\rho _{0}\cdot \exp (%
\frac{1}{2\cdot n}\cdot \overline{\varphi })\text{ .}  \label{3.10}
\end{equation}

The dilaton field $\overline{\varphi }$ could be represented by means of $%
\rho $ in the form 
\begin{equation}
\overline{\varphi }=2\cdot n\cdot (\log \frac{\rho }{\rho _{0}})\text{ .}
\label{3.11}
\end{equation}

Therefore, the dilaton field $\overline{\varphi }$ takes the role of a mass
density scaling factor for the rest mass density of a test particle. This is
another physical interpretation as the interpretation used by other authors
as mass field, pure geometric gauge field and etc.

Since $\overline{\varphi }=-n\cdot \log (e/e_{0})=2n\cdot \log (\rho /\rho
_{0})$, a relation between $e$ and $\rho $ follows in the form 
\begin{equation}
\rho ^{2}\cdot e=\rho _{0}^{2}\cdot e_{0}=\text{ const. }=l_{p}^{2}\text{ ,}
\label{4.13}
\end{equation}
which is exactly the condition (b) of the definition for a free moving
spinless test particle.

\section{Conclusion}

In the present paper the conditions are found under which a space with
affine connections and metrics and especially a Weyl's space admit
shear-free and expansion-free non-null vector fields as velocities of flows
or of test particles. In a Weyl's space the vector fields appear as
conformal Killing vector fields. In such type of spaces only the rotation
velocity is not vanishing. This fact could be used as a theoretical basis
for models in continuous media mechanics and in the modern gravitational
theories, where a rotation velocity could play an important role. Further,
necessary and sufficient conditions are found under which a free spinless
test particle could move in spaces with affine connections and metrics on an
auto-parallel curve. In Weyl's spaces with Weyl's covector, constructed by
the use of a dilaton field, the dilaton field appears as a scaling factor
for the rest mass density as well as for the velocity of the test particle.
The last fact leads to a new physical interpretation of a dilaton field in
classical field theories over spaces with affine connections and metrics and
especially over Weyl's spaces. The obtained results could be used in
constructing local criteria for experimental check-up of the existence of
non-metricity and torsion in realistic models of space-time.

\end{document}